\documentclass[aip,graphicx,reprint]{revtex4-1}

\usepackage{microtype}
\usepackage{graphicx}
\usepackage{braket}
\usepackage{hyperref}
\usepackage{multirow}
\usepackage[dvipsnames]{xcolor}
\usepackage[normalem]{ulem}

\begin{document}

%\preprint{}

\title[Quantum simulation of \texorpdfstring{CO$_2$}{CO2} chemisorption in an amine-functionalized metal-organic framework]{Quantum simulation  of \texorpdfstring{CO$_2$}{CO2} chemisorption in an amine-functionalized metal-organic framework}

\author{Jonathan R. Owens}
    \email{jon.r.owens@gevernova.com}
\affiliation{GE Vernova Advanced Research Center, Niskayuna, NY, 12309, USA}

\author{Marwa H. Farag} 
\affiliation{ Quantum Algorithm Engineering, NVIDIA, Santa Clara, CA, 95051, USA}

\author{Pooja Rao} 
\affiliation{ Quantum Algorithm Engineering, NVIDIA, Santa Clara, CA, 95051, USA}

\author{Annarita Giani}
\affiliation{GE Vernova Advanced Research Center, Niskayuna, NY, 12309, USA}

\date{\today}

\begin{abstract}

We perform a series of calculations using simulated QPUs, accelerated by the NVIDIA CUDA-Q platform, focusing on a molecular analog of an amine-functionalized metal-organic framework (MOF),  a promising class of materials for CO$_2$ capture. The variational quantum eigensolver (VQE) technique is employed, utilizing both the unitary coupled-cluster method with singles and doubles (UCCSD) and adaptive ans\"atze  within active spaces extracted from the larger material system. We explore active spaces of (6e,6o), (8e,8o), and (10e,10o), corresponding to 12, 16, and 20 qubits, respectively, and simulate them using CUDA-Q's GPU-accelerated state-vector simulator.  Gate fusion is shown to decrease circuit evaluation time by 2-3$\times$, while parameter shift decreases the number of epochs required for variational convergence. The ADAPT-VQE method decreases both the number of epochs required for convergence and reduces the number of circuit parameters across all active spaces, at the cost of an increased number of circuit evaluations. Combining ADAPT-VQE with the one- and two-electron integrals from a classical CASSCF calculation recovers more correlation energy, at the cost of increased computational time. The CO$_2$ binding energy is computed, and we observe and discuss how just increasing the active space size can lead to uneven recovery of correlation energy, making the predicted binding energies variable, even positive (\textit{i.e.}, energetically unfavorable) in some instances. This can be partially remedied by using an alternative approach to computing the binding energy that more evenly spreads the active space. This work explores  the application of VQE to a novel material system using large-scale simulated QPUs and provides some insight into various important consideration when performing these types of calculations, ultimately highlighting the challenges of studying chemisorption on near-term quantum machines, simulated or not .

\end{abstract}


\maketitle

\section{Introduction and background}\label{sec:intro}

    \subsection{The need for quantum computing}\label{subsec:need-for-qc}

    Quantum computing holds significant promise for improving our ability to solve certain complex computational problems beyond the capabilities of classical methods. Applications span a wide range of fields, including optimization  \cite{abbas_challenges_2024}, machine learning and artificial intelligence \cite{wang_comprehensive_2024}, computational fluid dynamics,\cite{gaitan_finding_2020}, physics \cite{di_meglio_quantum_2024,ivanov_quantum_2023,ciavarella_quantum_2023}, materials science \cite{gujarati_quantum_2023}, and quantum chemistry \cite{cao_ab_2023,mcardle_quantum_2020,cao_quantum_2019}.
    Of the many potential applications, it is has been suggested that the largest gains in speed and accuracy will be in domains where simulations of quantum systems are involved, such as quantum physics and chemistry\cite{schleich_chemically_2024}.

    Traditional methods in quantum physics and chemistry attempt to solve the many-body Schr\"{o}dinger equation, or some analogue, for a system of many nuclei and electrons. Due to its high dimensionality and many-body nature, this problem is intractable in its full form, except for a few small, highly simplified cases. Many methods have emerged to address this difficulty by attempting to simplify the problem. The most popular such method is density-functional theory (DFT), which combines the fact that reducing the dimensionality of the problem by solving for the energy as a functional of the electron \textit{density}\cite{hohenberg_inhomogeneous_1964} and rewriting the fully interacting electron picture in terms of a fictional, non-interacting potential is theoretically an exact approach to solving the fully-interacting problem\cite{kohn_self-consistent_1965}. The challenge lies in the fact that the \textit{true} form of the potential, the so-called exchange-correlation functional, is unknown, and thus approximations or best-guesses must enter in any practical implementation of DFT. While DFT has proved successful in a large number of domains, there are known difficulties in applying DFT to some types of materials, specifically ones exhibiting strong electron-electron correlation. There are a zoo of possible exchange-correlation functionals\cite{martin_electronic_2020}, with each trying to address various shortcomings of other functionals at describing particular classes of materials and physical processes.

    \subsection{\texorpdfstring{CO$_2$}{CO2} sorbent materials}\label{subsec:cc-sorbent-materials}

    While currently a very active area of research, the adsorption of CO$_2$ has been studied and practiced for some decades. Liquid-based amine sorbents, which capture CO$_2$ via the nitrogen groups to form carbamates\cite{miller_unraveling_2020}, are the most mature technology. They have not seen widespread adoption, however, for a few reasons: the liquid solvents are corrosive, prone to degradation, and require large amounts of energy to regenerate the sorbent by releasing CO$_2$\cite{reynolds_towards_2012}. Solid CO$_2$ sorbents, porous materials that can capture CO$_2$ via a physical (van der Waals), chemical, or combined\cite{zhu_high-capacity_2024} interaction have emerged as a promising alternative to liquid amines. A number of such material classes exist, including zeolites\cite{boer_zeolites_2023}, covalent organic frameworks\cite{li_covalent_2023}, alumina\cite{priyadarshini_direct_2023}, silicas\cite{sardo_unravelling_2024}, and metal-organic frameworks (MOFs)\cite{millward_metalorganic_2005, sumida_carbon_2012}. MOFs are materials that have metal nodes connected via organic linkers, forming (generally) 3-dimensional, porous, crystalline materials. They have been particularly popular in recent years, owing to the large search space generated from their structural and compositional tunability\cite{chung_advances_2019, moosavi_understanding_2020}, and the potential of solid sorbents to increase stability and reduce energy consumption upon regeneration\cite{millward_metalorganic_2005}. 

    MOFs are not without their shortcomings, however. A key challenge is that, in most practical applications, CO$_2$ is competing with other molecular species, such as H$_2$O, which can decrease the amount of CO$_2$ captured in practice. The introduction of amine groups into solid supports, like MOFs, has been explored as a way to retain the lower energetics and stability of solid sorbents, while retaining the selectivity of liquid amines \cite{mcdonald_capture_2012, forse_elucidating_2018, kim_cooperative_2020, milner_diaminopropane-appended_2017, sardo_unravelling_2024, wang_direct_2023, liu_dynamic_2010, bali_aminosilanes_2014,moon_underlying_2024}. Amine-functionalized MOFs, and similar materials, are ripe for large-scale computational studies. But the predictive accuracy of such calculations can be lacking, as we will discuss in Section \ref{subsec:qc-cc-sorbents}.

    \subsection{Quantum computing and \texorpdfstring{CO$_2$}{CO2} sorbents}\label{subsec:qc-cc-sorbents}

    In the context of carbon capture, in which sorbent materials are being developed to selectively capture carbon dioxide from a gas mixtures, such as the atmosphere or post-combustion flue gas, DFT has been routinely applied. DFT has played a role in screening large classes of materials, such as MOFs\cite{rosen_identifying_2019}, as well as providing detailed accounts of the chemical and physical processes occurring during the adsorption of CO$_2$\cite{zhang_unveiling_2020, damas_understanding_2021, huynh_theoretical_2023}. It is well-understood, however, that the choice of the exchange correlation functional and dispersion correction will affect the predicted energetics of both chemisorption\cite{lee_assessment_2022} and physisorption \cite{vlaisavljevich_performance_2017, lee_small-molecule_2015}, as well as the geometry of the predicted structure itself\cite{owens_understanding_2024}.

    Because of this, and the general interest for developing carbon capture materials, quantum computing has been suggested and explored as a possible solution to these challenges\cite{greene-diniz_modelling_2022, barroca_exploring_2024}. However, the size of existing quantum hardware is well-below what is needed to solve problems of practical scientific and industrial interest\cite{von_burg_quantum_2021}.

    In this work, we perform studies to assess how existing \textit{simulated} QPUs can scale and perform on a particular system of interest from the CO$_2$ capture literature, an amine-functionalized variant of Mg$_2$(dobpdc)\cite{siegelman_water_2019}. We apply state-of-the-art techniques in simulated QPUs, including GPU-acceleration and gate fusion using CUDA-Q platform\cite{The_CUDA-Q_development, cuquantum-paper}. As far as the authors are aware, VQE has not been applied to the \textit{chemisorption} of CO$_2$.

\section{Workflow}

\begin{figure*}
    \centering
    \includegraphics[width=\linewidth]{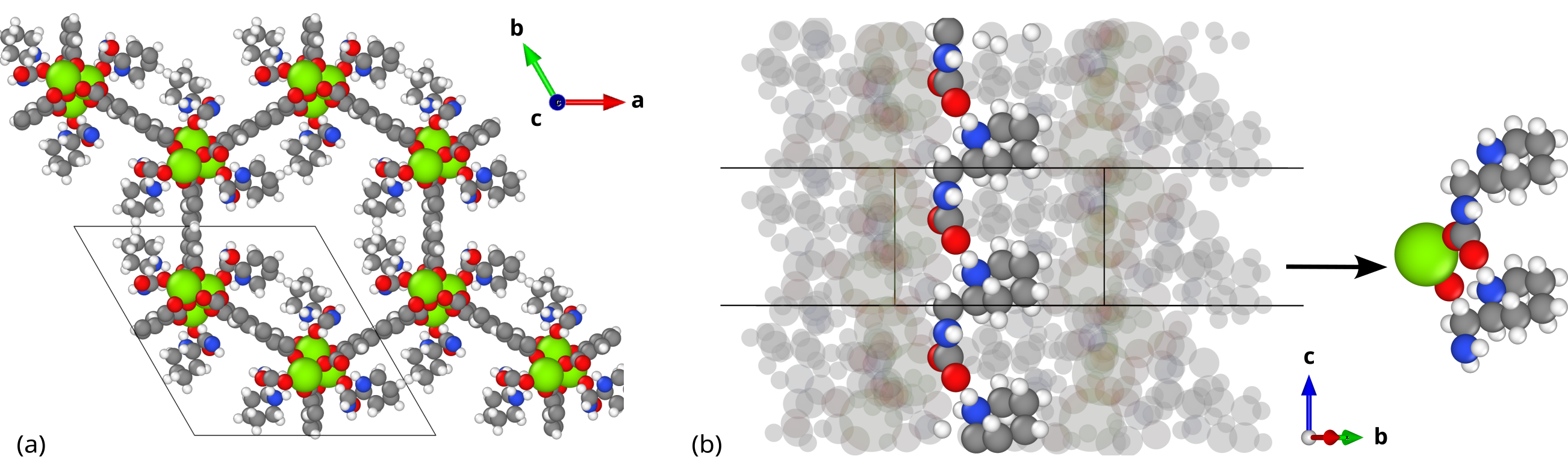}
    \caption{ (a) View of the ampd amine-functionalized metal-organic framework Mg$_2$(dobpdc) looking down the pore direction. The black lines mark the boundaries of one unit cell, with a super cell shown to illustrate the pore structure. (b) A view along the crystallographic \textit{c}-axis, showing the ammonium carbamate chain formation, mapped to the proxy chain model. Gray atoms are carbon, green are magnesium, blue are nitrogen, red are oxygen, and white are hydrogen.}
    \label{fig:periodic_to_molecule}
\end{figure*}

To perform quantum chemical calculations on simulated QPUs, a number of methods must be developed and integrated. The general workflow can be divided into three conceptual portions:

\begin{enumerate}
    \item Mapping the periodic structure to a representative gas-phase version. This is necessary to shrink the structure size to something manageable on a simulated quantum computer and to be compatible with our non-periodic code-base, while capturing as much of the relevant physics and chemistry as possible. The chosen molecular analog is shown in Fig. \ref{fig:periodic_to_molecule}(b). This mapping preserves the ammonium carbamate formation between adjacent amines down the crystallographic \textit{c}-axis. The rationale for this specific choice is discussed in Section \ref{subsec:mapping}
    \item Classical pre-processing and active space determination (Section \ref{subsec:preproc}.) This step involves choosing an appropriate basis and determining the orbitals that are likely to drive the chemistry of the system. 
    \item Variational quantum eigensolver (VQE) on a simulated QPU (Section \ref{subsec:vqe}.) This is an iterative, quantum-classical variational algorithm, in which a quantum device prepares a parameterized wavefunction $\ket{\Psi(\theta)}$, and the classical device optimizes those parameters, $\theta$. We search for the optimal $\theta$ of the UCCSD ans\"atz using different optimizers, on GPU-accelerated, simulated QPUs, ranging from 12 to 20 qubits. We additionally explore the use of an adaptive ans\"atz, ADAPT-VQE, with simulated QPUs of size 8 to 20 qubits.
\end{enumerate}

    \subsection{Mapping the periodic structure to a representative gas-phase version} \label{subsec:mapping}

    MOFs are typically 3-dimensional, periodic structures in all directions, with the number of atoms in the unit cell (without amine functionalization) on the order of 100. The insertion of amines increases the number of atoms in a unit cell even further. Our material of choice, 2-ampd-Mg$_2$(dobpdc), has 216 atoms in the unit cell, which has dimensions $|a| = |b| = 21.95$ \AA, $|c| = 7.01$ \AA. This structure is visualized in Figure \ref{fig:periodic_to_molecule}(a). Note the hexagonal pore shape and how the amines tend to pair up upon CO$_2$ adsorption.

    The molecular analog studied is shown in Figure \ref{fig:periodic_to_molecule}(b). This structure attempts to capture some of the physics of ammonium carbamate chain formation, keeping two adjacent amines in the crystallographic \textit{c}-direction, the magnesium node, and an backbone oxygen atom that acts as a hydrogen bond acceptor for the secondary amine. This choice reduces the structure to 49 atoms from the original 216.

    \subsection{Classical pre-processing and active space determination} \label{subsec:preproc}

    The gas-phase model, discussed in Section \ref{subsec:mapping}, has 49 atoms, 170 electrons, and 249 orbitals in the 6-31G basis. This is impossible to study using a full-configuration-interaction (FCI) approach on traditional processors, and all the electrons cannot be treated on current quantum hardware, given the low logical qubit count of contemporary machines. The problem must be decomposed into active and inactive orbitals. The inactive orbitals are determined by the size of the active space, specified at pre-processing time by setting the number of electrons ($N_{e,AS}$) and orbitals ($N_{o,AS}$) in the active space. For the VQE portion of the calculation, the number of qubits equal the number of spin natural orbitals in the active space, i.e., $N_{qubits} = N_{o,AS,\alpha} + N_{o,AS, \beta}$. The orbitals themselves come from the MP2 calculation.

\begin{figure*}
    \centering
    \includegraphics[width=\linewidth]{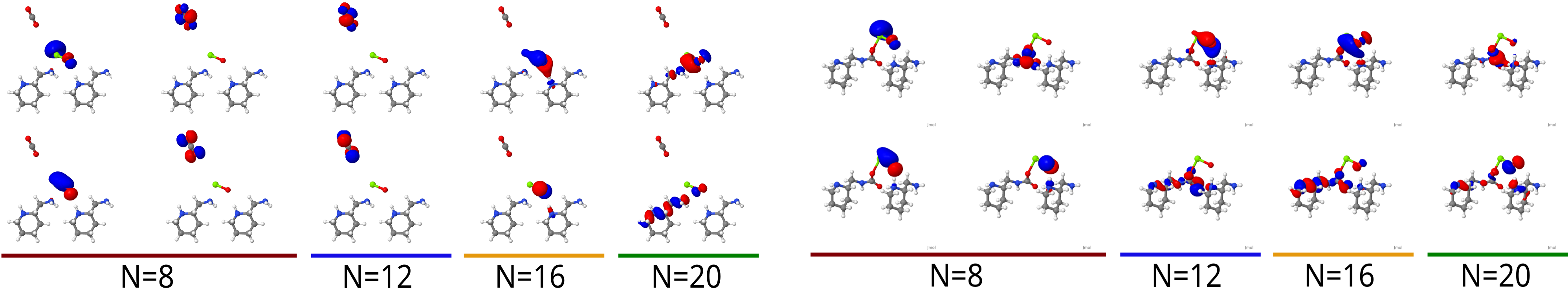}
    \caption{The natural orbitals treated in the active space in our simulations. The active space orbitals, computed from the MP2 theory, where chosen based on the natural orbital occupation number (NOON). The visualized orbitals on the left are for the system before the CO$_2$ has been adsorbed, and the orbitals on the right show the active space for the post-adsorbed structure. The colored lines indicate the orbitals included in simulations of the number of qubits $N$. In all cases, the active space consisted of an equal number of orbitals and electrons, meaning that the $N=8$ case had 4 orbitals and 4 electrons, $N=12$ had 6 orbitals and 6 electrons, and so on. The orbitals clearly correspond to chemically relevant regions of the adsorption process.  Gray atoms are carbon, green are magnesium, blue are nitrogen, red are oxygen, and white are hydrogen.}
    \label{fig:orbitals}
    \end{figure*}

    Visualizing the orbitals in the active space for our various simulation sizes will give some insight into the chemistry being studied as $N_{qubits}$ increases. The Moller-Plessent perturbation theory (MP2) natural orbitals for the active spaces studied in this work are visualized in Figure \ref{fig:orbitals}. These orbitals are determined during the classical pre-processing step. Inspection of Figure \ref{fig:orbitals} highlights that the selected active space orbitals are those around the CO$_2$ and the adsorption site (including the amine group and magnesium node).

    Based on the above active space definitions, we solve for the energy using classical methods, such as Hartree-Fock, Moller-Plessent perturbation theory (MP2), complete-active space self-consistent field (CASSCF), and coupled cluster with singles and doubles (CCSD), as implemented in PySCF. The energies in these various schemes allow us to have benchmark results against which to compare our VQE numbers. CASCI and CASSCF are used to get the one- and two-electron integrals for computing the molecular spin electronic Hamiltonian. These integrals are used to generate the spin Hamiltonian via the Jordan-Wigner mapping\cite{jordan_uber_1928}.

    \subsection{Computational Resources} \label{subsec:vqe}

    All simulations use the CUDA-Q platform\cite{The_CUDA-Q_development, cuquantum-paper, smaldone2024quantum} to implement the UCCSD-VQE\cite{CUDAQ_UCCSD-VQE} and ADAPT-VQE\cite{CUDAQ_ADAPT-VQE}. CUDA-Q is an open-source platform for integrating and programming QPUs, GPUs, and CPUs with a single system, with a GPU-accelerated state vector simulator. State vector simulations were run on the Perlmutter supercomputer at the National Energy Research Scientific Computing center (NERSC), equipped with NVIDIA A100 GPUs. Internal resources, comprised of NVIDIA H100 GPUs, at GE Vernova Advanced Research were used for additional calculations.

    \subsection{Algorithmic Considerations} \label{subsec:algo}

    Different strategies may be employed to make quantum simulations more tractable, particularly in the NISQ era or on QPU emulators. This section highlights gate fusion, parameter shift, and adaptive ans\"atze, three such strategies explored in this work.

    \subsubsection{Gate fusion} \label{subsubsec:gf}

\begin{figure}
        \includegraphics[width=0.48\textwidth]{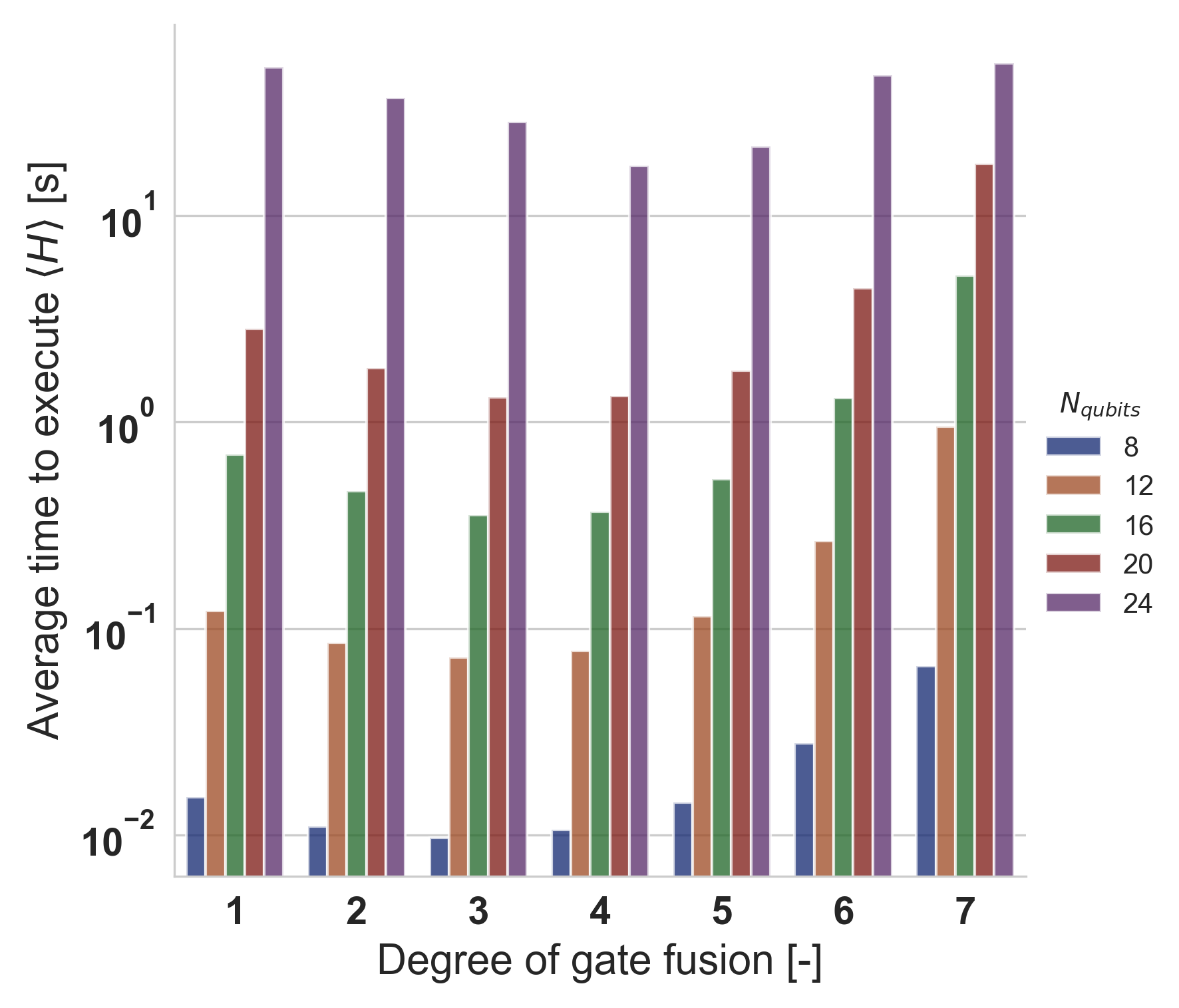}%
        \caption{
        The impact of the gate fusion hyperparameter, referred to as the degree of gate fusion, on the computation time of a single (averaged) expectation value call in CUDA-Q  for UCCSD-VQE. The optimal gate fusion degree is problem-dependent, with peak performance typically observed at degrees 3 to 4.
        }
        \label{fig:gate_fusion}
    \end{figure}

    Applying the sequence of unitary gates or matrices to the state vector is a time-intensive portion of a state vector-based QPU simulations. One method of speed up that reduces computational cost is gate fusion\cite{cuquantum-paper,cudaq-blog}, which is an available feature in CUDA-Q. Gate fusion is a process in which multiple quantum gates are combined into a single matrix. The degree of gate fusion is a hyperparameter in the UCCSD-VQE run, and it can greatly influence the runtime to evaluate a complete circuit to get $\braket{H}$, as shown in Figure \ref{fig:gate_fusion}. 
    
    Based on the plot, which has a log-scale on the \textit{y}-axis, we can see that gate fusion affords little practical speedup on the smaller active spaces. However, as the size of the active space increases, the wall-time speedup becomes appreciable. In fact, it is especially impactful given that the larger active spaces need more iterations, and thus $\braket{H}$, to converge. Without gate fusion, each $\braket{H}$ takes around 56 seconds. With the optimal degree of gate fusion, this can be reduced to around 20 seconds. This is almost about a 2.8x speedup, meaning that in the absence of such a technique, VQE runs would take substantially longer.

    \subsubsection{Parameter shift} \label{subsubsec:ps}

    The expectation value of a VQE circuit can be thought of as a quantum function, $f(\theta)$, parameterized by some $\theta = \theta_1, \theta_2, ...$. The exercise in a VQE calculation is to find the optimal $\theta_i$ parameters that minimize the value of $f$. Since the partial derivative of $f(\theta)$ can be expressed as a linear combination of other quantum functions that (typically) use the same circuit, shifted only by some offset, we can compute the partial derivatives of a variational circuit using the same circuit architecture. The parameter shift computes the gradient by evaluating $f(\theta)$ at the shifted value \(\theta \pm \frac{\pi}{4}\), avoiding the need for finite differences. Mitarai, \textit{et al.}\cite{mitarai_quantum_2018} and Schuld, \textit{et al.}\cite{schuld_evaluating_2019} discuss the rules in mathematical detail, and we refer the interested reader to their works. Paramater shift allows a straight-forward approach of computing the gradient of a circuit and thus can leverage a gradient-based optimizer to decrease the number of epochs needed to reach the minimum value of $f(\theta)$, as shown Section \ref{subsec:uccsd-vs-adapt}.

    \subsubsection{Adaptive ans\"atz}

    The unitary coupled-cluster with single and double excitations (UCCSD) has been one of the most popular chemistry-inspired ans\"atze studied in the literature. Due to its fixed form, it has a fixed number of parameters that must be optimized using the VQE, leading to difficulty in converging the energy as the size of the active space grows. ADAPT-VQE is an alternative technique, in which one pool of all the operators is selected each iteration, and therefore the wavefunction ansatz is built iteratively\cite{grimsley_adaptive_2019, mukherjee_comparative_2023, barroca_exploring_2024}. The pool of operators can be arbitrarily large, though it typically is defined as the UCCSD operators. The ans\"atz grows upon each iteration, adding the operator that most effectively lowers the energy. In this way, you can build a more compact ans\"atz, keeping only the required operators to minimize the energy and represent the wavefunction. This increases the convergence efficiency and the simulation's ability to saturate chemical accuracy, as previously demonstrated and shown in Section \ref{subsec:uccsd-vs-adapt}. The use of ADAPT-VQE essentially negates the need for gate fusion, as the circuit representation is much more efficient by its nature.

\section{Results} \label{sec:results}

\subsection{Comparing UCCSD-VQE and ADAPT-VQE} \label{subsec:uccsd-vs-adapt}

\begin{figure*}
    \centering
    \includegraphics[width=\linewidth]{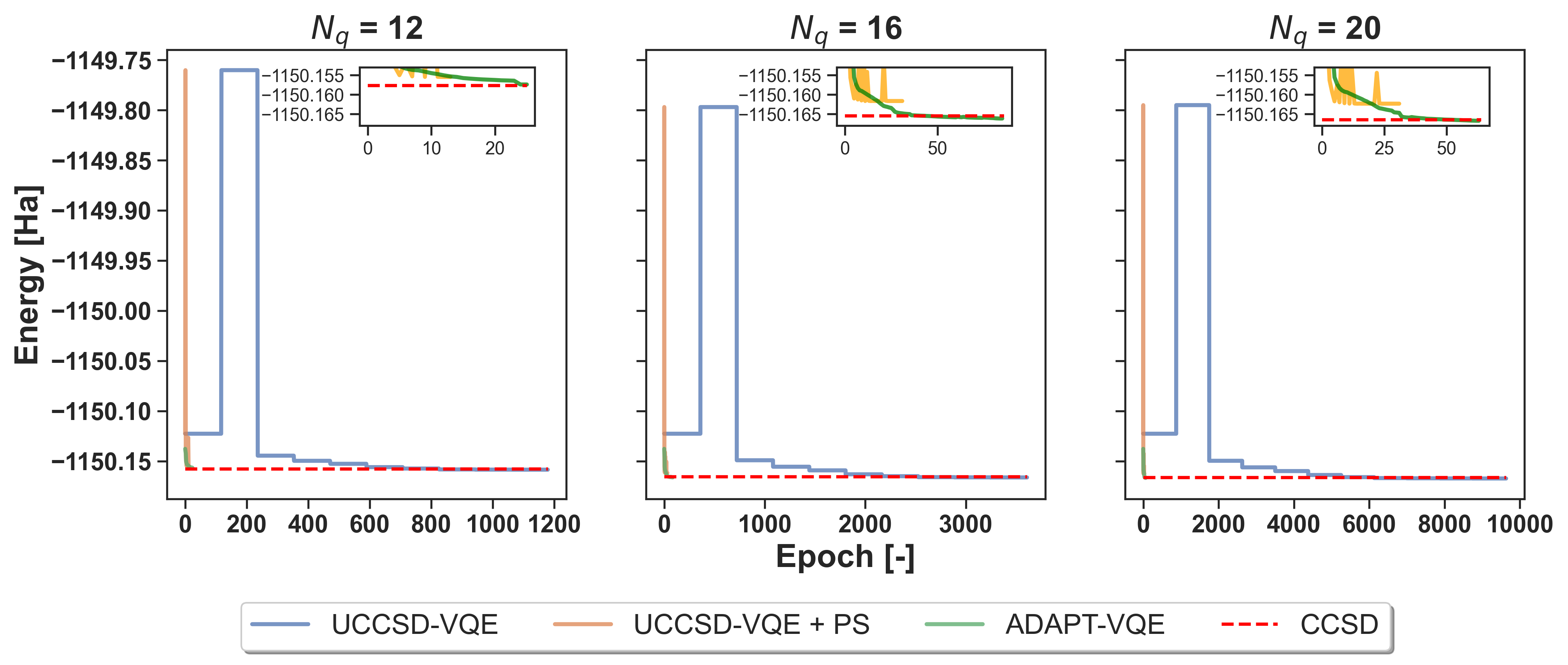}
    \caption{Energy vs. VQE epoch for 12, 16, and 20 qubit simulations. The \textit{y}-axes are the same for all the runs. The blue lines represent the UCCSD-VQE run, the orange lines represent the UCCSD-VQE runs with parameter shift used for gradient computation, the green lines represent the ADAPT-VQE approach, and the red dotted lines show the classical CCSD energies. The insets focus on the region where UCCSD-VQE+PS and ADAPT-VQE approach the classical result. In all three active space sizes, UCCSD-VQE and ADAPT-VQE reach chemical accuracy compared to CCSD, with ADAPT-VQE saturating the energy value much more rapidly. UCCSD-VQE+PS, however, does not quite converge to the classical CCSD results.}
    \label{fig:summary_results}
\end{figure*}

    This section is focused on evaluating the performance of a few different VQE schemes, comparing their performance with respect to accuracy, circuit complexity, and convergence. In particular, UCCSD-VQE (with and without parameter shift) and ADAPT-VQE are compared between themselves and classical CCSD results. The VQE-UCCSD (without parameter shift) runs used the limited-memory Broyden–Fletcher–Goldfarb–Shanno algorithm (L-BFGS-B), as implemented in SciPy\cite{virtanen_scipy_2020} as the optimizer.

    Figure \ref{fig:summary_results} compares the convergence of UCCSD-VQE, UCCSD-VQE with parameter shift, and ADAPT-VQE as a function of epoch, as compared to the classical CCSD results, for active spaces consisting of (6e,6o), (8e,8o), and (10e,10o). UCCSD-VQE converges past the CCSD reference value for all three active spaces considered. However, the number of epochs is large: 1180, 3610, and 9636 for the 12, 16, and 20 qubit runs, respectively. Using the parameter shift technique, the number of epochs is reduced respectively to 14, 32, and 32. Unfortunately, the parameter shift implementation does not saturate the CCSD reference value, even with a stricter convergence criteria. ADAPT-VQE presents a compromise in number of epochs between the two, while still converging below the CCSD reference value in 26, 86, and 64 epochs. While both ADAPT-VQE and UCCSD-VQE+PS have fewer total epochs, the number of circuit evaluations is higher, as shown in Table \ref{tab:vqe_comp}. The fact that ADAPT-VQE and UCCSD-VQE+PS have a similar number of circuit evaluations is intuitive, given that both approaches have to calculate the gradient of all the parameters. ADAPT-VQE provides a much more compact representation of the wavefunction, with the number of parameters in the ADAPT circuit being 26, 86, and 64 compared to 117, 360, and 875 for UCCSD. This compactification helps explain the reduced number of epochs required for convergence. It is interesting to note that the number of parameters in the adaptive ans\"atz is larger for the 16 qubit active space (86) than the 20 qubit one (64). One possible explanation is that the richer selection of orbitals in the larger active space (see Figure \ref{fig:orbitals}) provide a more targeted selection of the most relevant operators at each step. Recent work has indeed shown that active space choice can drive how efficient the circuit representation is\cite{vaquero-sabater_physically_2024}.

    Even though parameter shift and ADAPT-VQE reduce the number of epochs to variational convergence, they do have substantially more circuit evaluations than UCCSD-VQE, as shown in Table \ref{tab:vqe_comp}. This is because of the required gradients computed at each step for each parameter. This cost could potentially be offset by leveraging a multi-QPU approach, where these independent gradient evaluations could be farmed out to multiple QPUs without any substantial overhead.

    \begin{table}
        \resizebox{0.45\textwidth}{!}{%
        \centering
        \begin{tabular}{|c|c|c|c|}\hline
             & \textbf{12 qubits} & \textbf{16 qubits} & \textbf{20 qubits} \\ \hline
            \textbf{\#e:\#o} & 6:6 & 8:8 & 10:10 \\ \hline
            \multicolumn{1}{|c|}{\textbf{$E_{\text{HF}}$ [Ha]}} & \multicolumn{3}{|c|}{-1150.123} \\ \hline
            \textbf{$E_{\text{CCSD}}$ [Ha]} & -1150.158 & -1150.165 & -1150.167  \\ \hline 
            \textbf{$E_{\text{UCCSD-VQE}}$ [Ha]} & -1150.159 & -1150.166 & -1150.167  \\ \hline 
            \textbf{$E_{\text{ADAPT-VQE}}$ [Ha]} & -1150.158 & -1150.166 & -1150.167  \\ \hline 
            \textbf{\# UCCSD params} & 117 & 360 & 875  \\ \hline 
            \textbf{\# ADAPT params} & 26 & 86 & 64  \\ \hline 
            \textbf{UCCSD circuit evals.} & 1180 & 3610 & 9636  \\ \hline 
            \textbf{UCCSD+PS circuit evals.} & 3042 & 30960 & 56032  \\ \hline 
            \textbf{ADAPT circuit evals.} & 3016 & 30874 & 55936  \\ \hline 
        \end{tabular}}
        \caption{Comparison table of the results for the 12, 16, and 20-qubit runs, comparing UCCSD-VQE (with and without parameter shift) and ADAPT-VQE. for a VQE run with and without parameter shift. The HF and MP2 energies are independent of active space sizes, which is why they span all of the active space columns.}
        \label{tab:vqe_comp}
    \end{table}
    
    \subsection{Computing the binding energy of CO$_2$} \label{subsec:chain-model}

    The binding energy of CO$_2$ in a MOF is generally computed by:

    \begin{equation}\label{eq:be}
        \Delta E_{\text{b}} = E_{\text{MOF + CO$_2$}} - E_{\text{MOF}} - E_{\text{CO$_2$}}.
    \end{equation}

    \noindent This requires additional quantum simulation of two systems: one similar to the structure examined in Section \ref{subsec:uccsd-vs-adapt} and gaseous CO$_2$. In this Section, calculations like the one in the previous section are run for these additional two systems, as well as a few additional simulations on all three systems so that the binding energy can be computed.

    \begin{figure}
    \centering
    \includegraphics[width=0.48\textwidth]{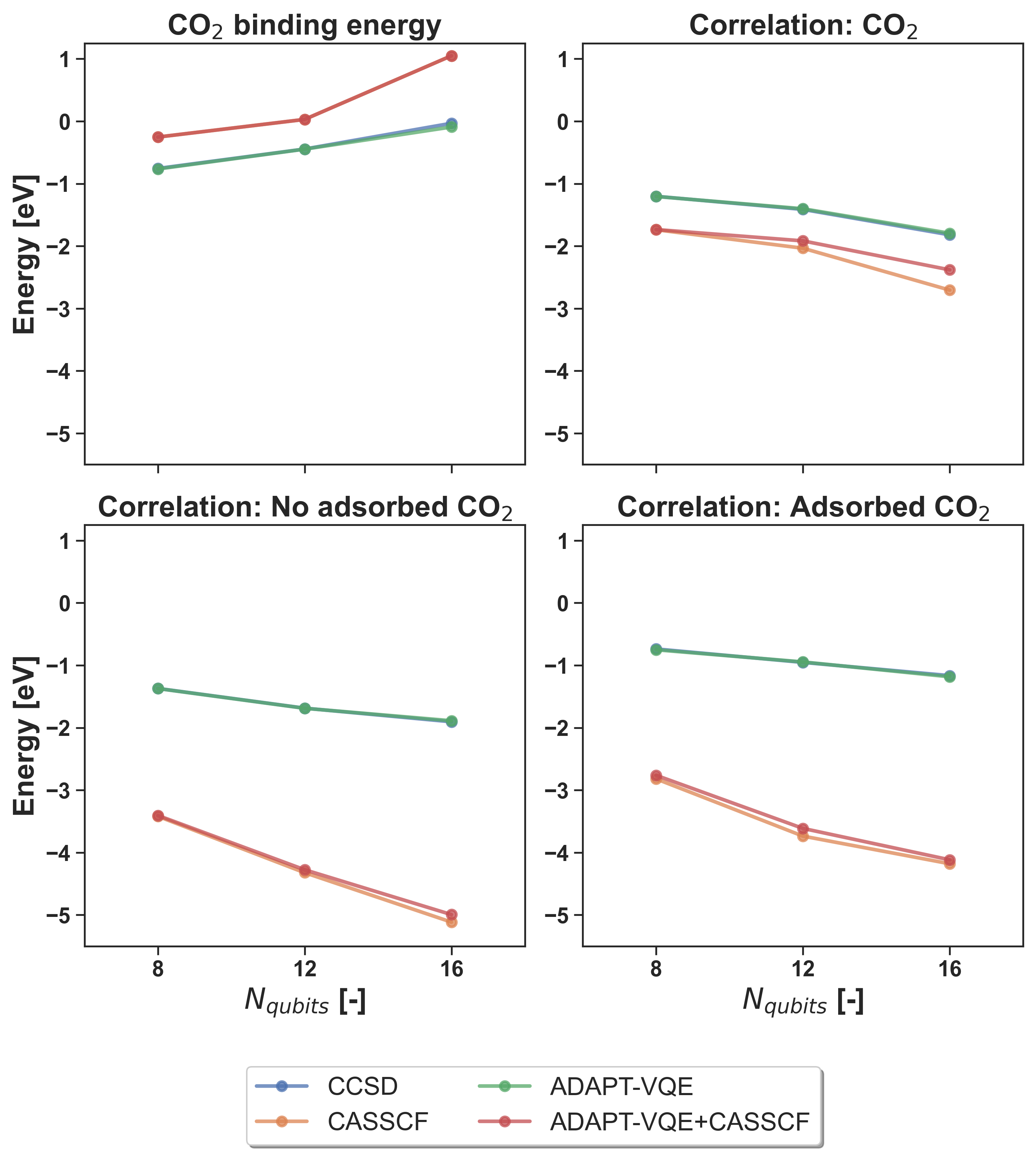}
    \caption{The CO$_2$ binding energy for different active space sizes, as well as the amount of correlation recovered in the different terms that go into computing the binding energy, as defined by eq. \ref{eq:be}. Increasing the active space size }
    \label{fig:standard-be-results}
    \end{figure}

    Figure \ref{fig:standard-be-results} shows the results of the CO$_2$ binding energy, as well as the amount of correlation recovered, as a function of active space size. The most important quantity, the CO$_2$ binding energy, is shown in the top left panel. Experimental measurements of the binding energy for the full ampd+Mg$_2$(dobpdc) material are around -0.7 eV\cite{siegelman_water_2019}. Interestingly, the quality of the predicted binding energy gets worse as the active space size is increased, with the values becoming energetically \textit{unfavorable} in simulations with 12 and 16 qubits. Given this trend, for these types of calculations, active spaces of 20 qubits were not simulated.

    Quantifying the correlation energy for each of the components that go into the binding energy helps in understanding the observation. The first set of calculations we performed were CCSD and ADAPT-VQE. For gaseous CO$_2$, increasing the active space sizes very quickly recovers the correlation energy. This is logical, as CO$_2$ is a small molecule and the larger active space sizes quickly approach covering all orbitals present in CO$_2$. The second largest system, the chain model with no adsorbed CO$_2$ (bottom left panel in Fig. \ref{fig:standard-be-results}) recovers correlation energy relatively quickly, as well, when compared to the chain model with adsorbed CO$_2$ (bottom right panel in Fig. \ref{fig:standard-be-results}). Because the increase of the correlation for the pre-adsorbed components (right two terms in Eq. \ref{eq:be}) outpaces that of the adsorbed system (first term in Eq. \ref{eq:be}), it makes sense that the the predicted binding energy would become less favorable.

    \begin{figure*}
    \centering
    \includegraphics[width=\linewidth]{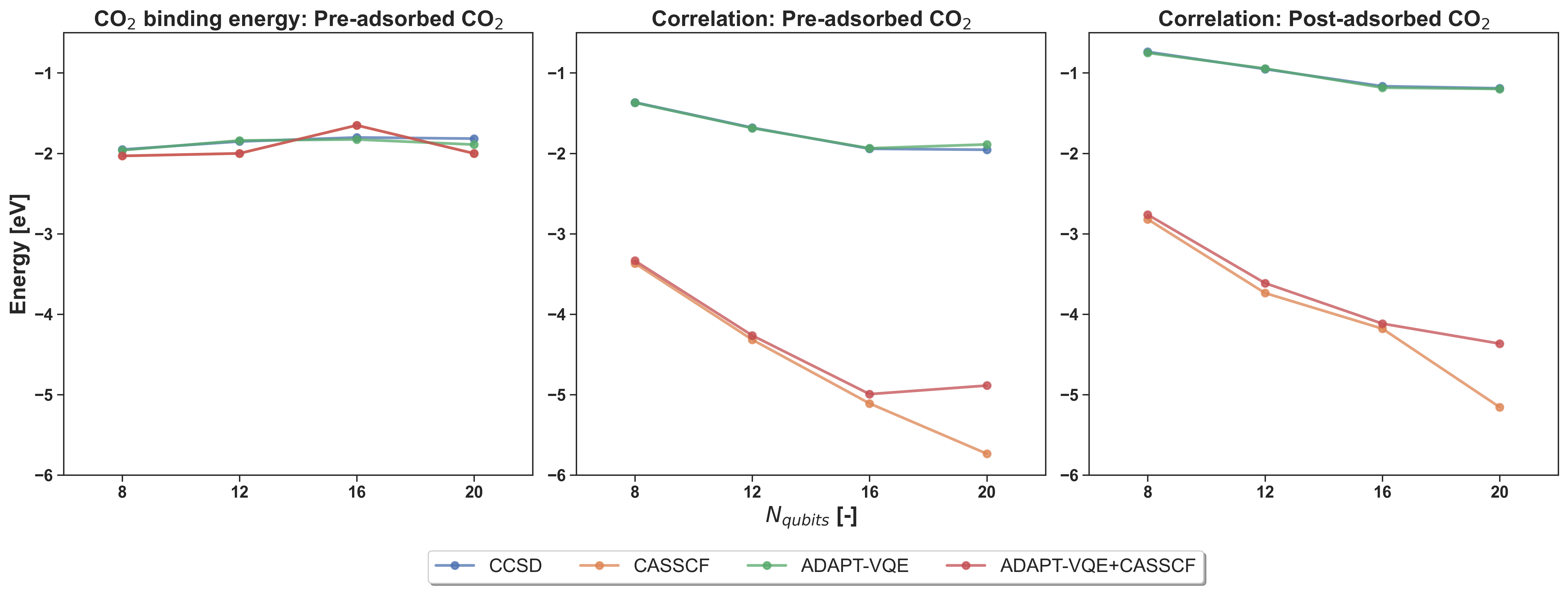}
    \caption{Binding and correlation energies of the alternative form of computing the binding energy, as defined in eq. \ref{eq:be_revised}. In this case, the binding energies are more physically meaningful, as they actually show energetic favorability, even at larger active spaces. The binding energy is not particularly sensitive to these larger active spaces, however, even though they recover substantially more correlation, especially when using ADAPT-VQE+CASSCF. This is because of the cancellation of this additional correlation, as it trends very similarly, as seen the right two plots above.}
    \label{fig:interacting-results}
    \end{figure*}

    In general, choosing the correct active space can make a large difference in the amount of correlation recovered. Recent work has shown the value of CASSCF's ability to recover additional correlation, as it also optimizes the orbitals\cite{de_gracia_trivino_complete_2023}. Orbital correlation can be significant, especially in larger molecules. Recovering more of the correlation energy for the larger molecular systems may improve the binding energy. Additionally, it is clear from the energies in Table \ref{tab:vqe_comp} that there is not actually that much correlation recovered with the UCCSD-type methods as the active space size increases: -1150.58 Ha in the 12 qubit case to -1150.167 Ha in the 20 qubit case, which is around 0.3 eV. CASSCF's ability to deal better with static correlation than CCSD may improve the situation.
    
    To explore this idea in more detail for this application, the one- and two-electron integrals from a classical CASSCF calculation are used in the ADAPT-VQE calculation, denoted ADAPT-VQE+CASSCF. Based on the plots in Fig. \ref{fig:standard-be-results}, it is clear that more correlation energy is recovered: the slope of the ADAPT-VQE+CASSCF is steeper as a function of active space size. However, while the amount of correlation energy increases in the larger molecules, the correlation recovered in the model without CO$_2$ still outstrips the material with CO$_2$, meaning that the quality of binding energies still degrades, instead of improving.

    A potential workaround to this issue is to alter the way the binding energy is computed. Instead of Eq. \ref{eq:be}, one can generate a structure that has CO$_2$ present but not adsorbed. The new binding energy can be computed as:

    \begin{equation}\label{eq:be_revised}
        \Delta E_{\text{b}} = E_{\text{MOF + CO$_2$,post-ads}} - E_{\text{MOF + CO$_2$,pre-ads}}.
    \end{equation}

    \noindent In this situation, both the pre- and post-adsorbed structures have the same number of atoms, electrons, and orbitals. The active space will thus be more evenly spread throughout the system, with the active orbitals chosen based on the occupation numbers, as compared both to the adsorbent and adsorbate. This approach is similar to how binding energies can be computed when doing reaction barrier calculations with the nudged-elastic band (NEB) method. A structure is constructed with the adsorbate far enough away from the adsorption site to still be considered ``free''. During the course of the calculation, the adsorbate moves closer and eventually adsorbs, with the final state being the energy of the adsorbed configuration. The binding energy in these calculations is the difference between the initial and final configurations.
    
    Using this method, the CO$_2$ binding energies and correlation energies for the different calculation schemes are shown in Fig. \ref{fig:interacting-results}. With this updated framework, the binding energies still increase as the size of the active space does, as they did in the results shown in in Fig. \ref{fig:standard-be-results}. However, at -2 eV, the initial value of the binding energy was far too strong for CO$_2$ adsorption in this system. The increasing active space size thus improves the binding energy relative to the very negative initial values. Another reason the binding values may be so negative is the fact that the structures have not undergone energy minimization via relaxing the geometries. This typical step of the workflow has been omitted because the molecularized structure is too isolated and is missing structural components that would keep it together, though the structures were initially built from the fully-relaxed periodic system at the vdW-DF2 level\cite{siegelman_water_2019}, in an attempt to preserve a low-energy electronic environment around the CO$_2$ adsorption.  This approach is not without precedent\cite{greene-diniz_modelling_2022}, but it is a challenge when using a proxy model like this one. These issues are likely exacerbated when studying chemisorption, because the tendency of the orbitals relaxing to the state is sensitive to the local structural environment.

    \subsection{Isolated amp-d} \label{subsec:gas-ampd}

    Studying two gaseous amp-d molecules reacting with CO$_2$ to form a carbamate species may be more suitable for near-term quantum simulation. While the number of the atoms in the system is around the same as the proxy model for the periodic system (47 vs. 49 atoms), the structure itself is more amenable to a robust computational workflow. Whereas the periodic model was constructed through removal of the surrounding atoms, keeping only the ones deemed most chemically relevant, the isolated amine system can be fully relaxed through DFT calculations, identifying the lowest structural configuration upon adsorption. These fully minimized structures are then used as inputs to quantum simulation (or classical active space methods). 
    
    As the quantum calculations generally showed similar behavior as the classical quantum chemistry calculations (Figures \ref{fig:standard-be-results} and \ref{fig:interacting-results}), we performed CCSD and CASSCF calculations on the gaseous system to see if it may be more amenable to quantum simulation. We observe the same trends as reported in Section \ref{subsec:chain-model}. For one, the difference in the recovery rate of the correlation energy between the different systems makes direct calculation of the binding energy with Eq. \ref{eq:be} face the same issues as with the molecularized model. As before, using a pre- and post-adsorbed state gives more physically meaningful binding energies. However, even extending to active spaces of size (16e,16o) shows no obvious signs of the binding energies or correlation energies converging, as shown in Figure S1, indicating even larger active space sizes may be required to approach chemical accuracy, or additional methods of more balanced correlation energy recovery need to be explored.

\section{Discussion} \label{sec:discussion}

Initial calculations of a molecularized analog of ampd-Mg$_2$(dobpdc) using UCCSD-VQE (with and without parameter shift) and ADAPT-VQE on chemically relevant active spaces, spanning from (6e,6o) to (10e,10o) showed that the ADAPT-VQE approach provides a more efficient ans\"atz and has fewer epochs to convergence than UCCSD-VQE, while retaining the chemical accuracy compared to the classical CCSD. While the parameter shift approach does result in the lowest number of epochs to convergence, it does not reach the same level of accuracy as the adaptive approach. If one is interested in the more traditional UCCSD-VQE scheme, a combination of parameter shift (to reduce convergence time) as a starting point of the calculation could be later continued with a more traditional optimizer. Additionally, gate fusion provides and alternative approach optimizing the circuit and reducing the type to execute $\braket{H}$.

The adaptive VQE allowed the calculation of the binding energy of CO$_2$ according to the standard approach, defined in eq. \ref{eq:be}. The biased nature of correlation recovery in the smaller molecules (CO$_2$ itself and the amine complex without CO$_2$ adsorbed) meant that the quality of the binding energy actually decreased as the active space size increased. While using the CASSCF integrals (ADAPT-VQE+CASSCF) indeed recovered more correlation energy than the more standard approach, the recovery was still uneven, leading to positive predicted binding energies in the (8e,8o) active space. Being especially specific about the orbitals considered with an approach like the automated construction of molecular active spaces (AVAS) from atomic valence orbitals\cite{sayfutyarova_automated_2017} may provide finer-grain control over the chemistry studied. However, there are too many chemically relevant orbitals (e.g., Mg-$3s$, N-$2s$, N-$2p$, O-$2s$, O-$2p$, etc.) in our studied system to fit on near-term quantum devices.

As an alternative approach, intended to balance the spread of the recovered correlation, we considered computing the binding energy more akin to the way used in NEB calculations (eq. \ref{eq:be_revised}). This did improve the results, as the binding energies retained more physically meaningful values (i.e., were negative), although the system bound much more strongly than one would expect compared to experiment (-2 eV vs -0.7 eV). The ADAPT-VQE+CASSCF method recovered substantial correlation as the active space size increased, especially when compared to the more traditional ADAPT-VQE approach (see Figure \ref{fig:interacting-results}). Continued work in this space could use approaches specifically developed to more evenly distribute the correlation recovery, such as ACE-of-SPADE\cite{kolodzeiski_automated_2023}.

Overall, the ADAPT-VQE+CASSCF method shows the most promise in the studied systems when considering the amount of correlation energy recovered, specifically as the active space size increases. As implemented, however, this approach suffers from the same scaling issue as CASSCF: the integrals from a classical CASSCF calculation are used as inputs to the ADAPT-VQE simulation, meaning that the approach becomes infeasible in its current form. de Gracia Triviño, \textit{et al.} showed that this can be bypassed by implementing a \textit{quantum} CASSCF scheme that would allow reclaiming of the scaling favorability of quantum algorithms\cite{de_gracia_trivino_complete_2023}. It is also worth pointing out that the ADAPT-VQE+CASSCF approach does necessitate less-compact circuits compared to typical ADAPT-VQE, meaning that circuit complexity could be an issue for this method until larger hardware is available. 

Scaling approaches like ADAPT-VQE+CASSCF to larger active spaces, as the appropriate hardware becomes available, may indeed allow computing physical quantities like binding energies to a higher-accuracy than existing approaches, so long as the CASSCF calculations can be offloaded to a quantum device. While the work here investigated ADAPT-VQE+CASSCF in a molecular system, the same approach could be used in a periodic system, increasing the applicability of the techniques without requiring compromises on structure size and the physics studied.

Alternatively, recent work proposed the generative quantum eigensolver (GQE)\cite{nakaji_generative_2024} to address some VQE shortcomings. Although this approach of using generative AI to generate a quantum circuit is promising, more work is needed to improve the GQE algorithm to calculate the electronic energies within chemical accuracy. 

Energy convergence is particularly important when computing binding energies, especially of small adsorbates, like CO$_2$. Inspection of Figures \ref{fig:standard-be-results} and \ref{fig:interacting-results} demonstrates that the magnitude of the binding energy can be 4-5x smaller than the correlation energy. Fortuitous cancellation of errors may remedy this, although that is not necessarily true, as our results show. As typical binding energies can be around -0.5 eV, chemical accuracy is important. Problems may even arise if the VQE solver has some discrepancy between the classical approach (CCSD or CASSCF), as those energy deltas could easily be around the same size as the binding energy. This challenge is not limited to quantum simulation methods. Indeed, poorly converged DFT calculations or using structures that are in local minima can present the same problems. In any scenario, care must be taken when computing binding energies of materials like the ones studied in this work.

While our mapping of the periodic system to a gaseous analog attempted to capture the relevant physics, any simplification will lost some important energetic terms. Even still, simulating a large super-cell to mimic the bulk materials of our system will require a large number of logical qubits (thousands) that will not be accessible in the near term quantum devices. However, methods like density-matrix embedding theory\cite{battaglia_general_2024, wouters_practical_2016, mitra_periodic_2022} or circuit knitting\cite{piveteau_circuit_2024}, combined with the considerations discussed in this paper, could allow simulation of such a large system on near term quantum devices.

Studying gas phase reactions may be more amenable to near-term quantum simulation, as discussed in Section \ref{subsec:gas-ampd}. Still, the binding energy is a challenging quantity for these methods to capture accurately, with particular care needed to avoid uneven recovery of the correlation energy across different molecules. Our binding energy calculations, for both the gaseous and periodic-analog models indicate that applying VQE-type active space methods to the study of CO$_2$ chemisorption will face challenges in balancing correlation energy recovery using active space sizes accessible in the near-term.

\section{Conclusion}

We used a GPU-accelerated state-vector simulator available in CUDA-Q to study ampd-Mg$_2$(dobpdc), a promising chemisorbent material for CO$_2$, highlighting a new application of the VQE simulations to chemisorption in amine-functionalized metal-organic frameworks. Gate fusion allows a reduction in the execution time of $\braket{H}$, especially as the active space size grows. Parameter shift and ADAPT-VQE allow for faster convergence of the variational algorithms, as compared with the standard UCCSD-VQE approach. ADAPT-VQE further allows a more compact circuit with fewer parameters. Computing the binding energy of CO$_2$ in a molecularized version of ampd-Mg$_2$(dobpdc) shows some challenge with active space scaling, which can be partially remedied by more even distribution of the active orbitals and the use of more robust methods, such as ADAPT-VQE+CASSCF. To enable scaling to even larger active space sizes beyond 20 qubits, implementation of approaches such as quantum CASSCF\cite{de_gracia_trivino_complete_2023} should be considered. Methods such as ACE-of-SPADE\cite{battaglia_general_2024} are candidates for even more robust active space selections in binding energy calculations, where even-handed distribution of the active space is required.  These strategies must be combined with methods more suitable to the periodicity of the system, such as DMET or circuit knitting.  Alternatively, the study of adsorption in a true molecular amine system may be an alternative system of study for near term quantum hardware, though we observed similar challenges as with the molecularized version of the periodic system. All of these approaches will be fruitful direction for further research.

\section*{Supplementary Material}

CO2 binding energies for two interacting ampd molecules in the gas phase computed with classical CCSD and CASSCF, along with a visualization of the structures and explanation of the process used for computing the binding energies.

\section*{Data Availability Statement}

The data that supports the findings of this study are available within the article. CUDA-Q implementations of UCCSD-VQE and ADAPT-VQE are available in the \href{https://github.com/NVIDIA/cuda-quantum/tree/main/docs/sphinx/applications/python}{CUDA-Q repository}. Some convenience scripts for batch-processing different systems are available \href{https://github.com/jonathan-owens/quantum-mcp}{here}.

\begin{acknowledgments}
This research used resources of the National Energy Research Scientific Computing Center (NERSC), a U.S. Department of Energy Office of Science User Facility located at Lawrence Berkeley National Laboratory, operated under Contract No. DE-AC02-05CH11231 using NERSC Award No. ALCC-ERCAP0025949 and NERSC Award No. DDR-ERCAP0025715.
\end{acknowledgments}

%\end{acknowledgments}

\bibliography{references}

\end{document}